\documentclass[a4paper,12pt]{article}
\usepackage{amssymb,float,amsmath,amsfonts,amsthm,cases,cite}
\usepackage[normalem]{ulem}

\usepackage[pdfencoding=auto]{hyperref}
\usepackage{xcolor}
\usepackage{hyperref}
\definecolor{linkcolor}{HTML}{799B03}
\definecolor{urlcolor}{HTML}{799B03}
\hypersetup{pdfstartview=FitH,linkcolor=linkcolor,urlcolor=urlcolor,colorlinks=true}

\begin{document}	
\begin{flushright}
INR-TH-2017-018
\end{flushright}

\vspace{10pt}

\begin{center}
{\LARGE \bf Perturbations in generalized Galileon theories}

\vspace{20pt}

R. Kolevatov$^{a,b}$\footnote[1]{\textbf{e-mail:} kolevatov@ms2.inr.ac.ru},
S. Mironov$^{a,c}$\footnote[2]{\textbf{e-mail:} sa.mironov\_1@physics.msu.ru},
V. Rubakov$^{a,b}$\footnote[3]{\textbf{e-mail:} rubakov@minus.inr.ac.ru}
N. Sukhov$^{b}$\footnote[4]{\textbf{e-mail:} nd.sukhov@physics.msu.ru},
V. Volkova$^{a,b}$\footnote[5]{\textbf{e-mail:} volkova.viktoriya@physics.msu.ru}

\vspace{15pt}

$^a$\textit{Institute for Nuclear Research of the Russian Academy of Sciences,\\
60th October Anniversary Prospect, 7a, 117312 Moscow, Russia}\\
\vspace{5pt}

$^b$\textit{Department of Particle Physics and Cosmology, Physics Faculty,\\
M.V. Lomonosov Moscow State University,\\
Vorobjevy Gory, 119991 Moscow, Russia}
	
$^c$\textit{Institute for Theoretical and Experimental Physics,\\
Bolshaya Cheriomyshkinskaya, 25, 117218 Moscow, Russia}
\end{center}

\vspace{5pt}

\begin{abstract}
We discuss the approaches by Deffayet et al. (DPSV) and Kobayashi et
al. (KYY) to the analysis of linearized scalar perturbations about
a spatially flat FLRW background in Horndeski theory. We identify
additional, potentially important terms in the DPSV approach. However
these terms vanish upon a judicious gauge choice. We derive a gauge
invariant quadratic action for metric and Galileon perturbations in
$\mathcal{L}_3$ and $\mathcal{L}_3+\mathcal{L}_4$ theories and show
that actions obtained in the DPSV and KYY approaches follow from this
gauge invariant action in particular gauges.
\end{abstract}

\section{Introduction and summary}
\label{sec:intro}
The theory of generalized Galileons (or, equivalently, Horndeski theory) has acquired 
significant interest lately~\cite{Horndeski:1974wa, Fairlie:1991qe, Luty:2003vm, Nicolis:2008in, Deffayet:2010zh, Deffayet:2010qz, Kobayashi:2011nu, Padilla:2012dx}, especially in the context of constructing cosmological models  with
the Null Energy Condition violation.
Galileons are scalar fields whose Lagrangians involve
second derivatives, but the corresponding field equations remain second order (for a review
see, e.g., Ref.~\cite{Rubakov:2014jja}). 

One of the purposes of utilizing Horndeski theories  is to construct bouncing and Genesis cosmological models free of instabilities (ghosts, gradient instabilities, etc.). While there are
examples of classical models that have no pathologies within a certain time period~\cite{Easson:2011zy, Cai:2012va,  Koehn:2013upa, Battarra:2014tga, Qiu:2015nha, Kobayashi:2015gga, Wan:2015hya, Ijjas:2016tpn}, the
majority of them have ``classical'' stability problems at late or early times.  Quite recently,
complete models, which are stable throughout the entire evolution, have been proposed
within the ``beyond Horndeski'' theory \cite{Cai:2016thi,Creminelli:2016zwa,Kolevatov:2017voe,Cai:2017dyi}.

In search of a complete bouncing or Genesis cosmological model one naturally starts
with the simplest  $\mathcal{L}_3$ subclass of Horndeski theory, 
with the Galileon field minimally coupled
to gravity. Its Lagrangian reads (mostly negative signature; $\kappa = 8\pi G$):
\begin{equation}
\label{eq:lagrangian}
\mathcal{L} = -\dfrac{1}{2\kappa}R + F(\pi,X) + K(\pi,X)\square\pi,
\end{equation}
where $\pi$ is the Galileon field, $F$ and $K$ are smooth functions, and
\begin{equation*}
X = g^{\mu\nu}\partial_\mu\pi\partial_\nu\pi,\quad \square\pi = g^{\mu\nu}\nabla_\mu\nabla_\nu\pi.
\end{equation*}
Generally, the issue of stability is determined by the highest momentum and frequency modes
of perturbations. Thus, second derivatives of perturbations in the equation of motion or,
equivalently, quadratic terms in the perturbed action are of main interest. 

In cubic Galileon theory, the danger of instabilities about cosmological backgrounds exists
in the scalar sector only. This sector has a single degree of freedom, which shows up differently
in different gauges. One way to study this sector at the quadratic level is to choose the unitary
gauge (the Galileon perturbations vanish, $\delta \pi = 0$), plug the perturbed metric  into the action,
expand the latter  to the second order and integrate out all non-dynamical degrees of freedom.
This approach was applied to the cubic theory  \eqref{eq:lagrangian} independently 
in Refs.~\cite{Deffayet:2010qz, Kobayashi:2010cm} and generalized to complete 
Horndeski theory in Ref.~\cite{Kobayashi:2011nu}. We refer to this approach 
as KYY for brevity.

Another method was proposed in Ref.~\cite{Deffayet:2010qz} (the DPSV approach). It was noted that the Galileon field equation and Einstein equations for the Lagrangian \eqref{eq:lagrangian} contain second derivatives
of both the metric and Galileon, and so do the linearized equations. The suggested
trick was to eliminate the second derivatives of the metric in the Galileon field equation by
employing the Einstein equations.  As a result, one obtains the equation which contains the
second derivatives of the Galileon only. Making use of this equation, one then restores the
derivative part of the quadratic action for the single scalar degree of freedom, which in this
approach is the Galileon perturbation. It should be noted that unlike the KYY approach, the
trick does not appear to require explicit gauge fixing.

While the KYY approach is quite general and widely used  for addressing the problem of stability~\cite{Kobayashi:2016xpl}, the DPSV trick was shown to be useful as an alternative~\cite{Libanov:2016kfc}. Moreover, the DPSV
approach significantly simplifies the study  of wormhole stability ~\cite{Rubakov:2016zah, Kolevatov:2016ppi,Evseev:2016ppw}.

Both methods are designed to obtain the derivative part of the quadratic action for
scalar perturbations, and hence the second derivative part of the unconstrained field equation.
There is a potential issue with DPSV formalism, however: though constraint variables (lapse
and shift perturbations $\delta N$, $\delta N_i$) enter the field equations without second derivatives, ignoring them 
(as was done originally~\cite{Deffayet:2010qz}) is dangerous,
as they are expressed, via constraints, through the
derivatives of the metric and Galileon. Therefore, the relationship between the KYY and DPSV
approaches is not entirely obvious. This is the issue we address in this paper. Our procedure
is to start with keeping both metric and Galileon perturbations in $\mathcal{L}_3$ theory and leaving
intact the most relevant gauge freedom. We then show that the results of the KYY and DPSV
approaches do correspond to specific gauge choices.

Another issue we address is whether the DPSV trick is applicable to more general subclasses
of Horndeski theory. We extend the method to the $\mathcal{L}_3+\mathcal{L}_4$ case. We find that the DPSV trick does not work for general backgrounds. However, in the case of a spatially flat cosmological background the DPSV approach applies and gives a correct description of scalar perturbations.

This paper is organized as follows. In Sec.~\ref{sec:theory} we obtain a gauge invariant quadratic
Lagrangian for metric and Galileon perturbations about a spatially flat FLRW background. In Sec.~\ref{sec:dpsv} we discuss the DPSV approach. As
we alluded  to above, we find that additional terms with metric perturbations have to be
taken into account in the equation of motion in comparison with the initial trick. However,
in the same section we show on fairly general grounds that there is a gauge choice which removes the new terms. The argument, however, may not apply to beyond Horndeski theories~\cite{Gleyzes:2014dya}.
In Sec.~\ref{sec:equiv} we give explicitly the gauges in which the gauge invariant Lagrangian for
metric and Galileon perturbations is reduced to KYY and DPSV  Lagrangians. In Sec.~\ref{sec:DPSV_for_L4}, the DPSV trick is applied to the $\mathcal{L}_3+\mathcal{L}_4$  subclass in a spatially flat FLRW background, and the results are shown to coincide with those obtained via  the KYY approach in Ref.~\cite{Kobayashi:2011nu}.

\section{Action for perturbations}
\label{sec:theory}
We study the theory in a spatially flat FLRW background:
\begin{equation*}
\mathrm{d}s^2 = \mathrm{d}t^2 - a^2 (t) \mathrm{d}{\bf x}^2, \qquad \pi = \pi (t) \; .
\end{equation*}
We begin with the quadratic action for scalar perturbations in cubic theory \eqref{eq:lagrangian}, which
contains metric and Galileon perturbations as well as their mixing. We fix the gauge only
partially, by setting the longitudinal part of the spatial metric equal to zero. So, the metric
perturbations read
\begin{equation}
\label{eq:metric_perturbations}
h_{00}=2\alpha,\quad h_{0i}=-\partial_i\beta,\quad h_{ij}=-a^2\cdot 2\zeta \delta_{ij},
\end{equation}
and we also consider Galileon perturbations $\chi$ about the homogeneous background:
\begin{equation*}
\pi \to \pi (t) + \chi.
\end{equation*}
We are interested in high momentum and frequency modes; therefore we neglect terms in
the action without derivatives of $\zeta$ and $\chi$, as well as those that are linear in their derivatives. On the other hand,
we keep all  terms that include $\alpha$ and $\partial_i\beta$. We comment on this issue later in the section. 

The second order action for metric and Galileon perturbations is
\begin{equation}
\label{eq:perturb_action_gr+gal}
\begin{aligned}
S^{(2)}_{gr+gal} = \int \mathrm{d}t\,\mathrm{d}^3x\,a^3 \Bigg (
&-\dfrac{3}{\kappa}\dot{\zeta}^2 + \dfrac{1}{\kappa}\dfrac{(\overrightarrow{\nabla}\zeta)^2}{a^2} + \Sigma \alpha^2 - 2\Theta\alpha\dfrac{\overrightarrow{\nabla}^2\beta}{a^2} + \dfrac{2}{\kappa}\dot{\zeta}\dfrac{\overrightarrow{\nabla}^2\beta}{a^2} + 6\Theta\alpha\dot{\zeta} - \dfrac{2}{\kappa}\alpha\dfrac{\overrightarrow{\nabla}^2\zeta}{a^2} \\
&+ 2\alpha\dfrac{\overrightarrow{\nabla}^2\chi}{a^2}K_X\dot{\pi}^2 - 2\dot{\chi}\dfrac{\overrightarrow{\nabla}^2\beta}{a^2}K_X\dot{\pi}^2 - 6\chi\ddot{\zeta}K_X\dot{\pi}^2 + 2\Gamma\alpha\dot{\chi} + 2\Lambda\dfrac{\overrightarrow{\nabla}^2\beta}{a^2}\chi\\
&+\mathcal{A}\dot{\chi}^2 - \mathcal{B}\dfrac{(\overrightarrow{\nabla}\chi)^2}{a^2}\Bigg),
\end{aligned}
\end{equation}
where
\begin{subequations}
\label{eq:notations_in_action}
\begin{align}
\mathcal{A} &= 2F_{XX}\dot{\pi}^2 + F_X - K_\pi - K_{\pi X}\dot{\pi}^2 + 6K_{XX}H\dot{\pi}^3 + 6K_XH\dot{\pi},\\
\mathcal{B} &=F_X - K_\pi + K_{\pi X}\dot{\pi}^2 + 2K_{XX}\ddot{\pi}\dot{\pi}^2 + 2K_X\ddot{\pi} + 4K_XH\dot{\pi},\\
 \Sigma &= \mathcal{A} \dot{\pi}^2 + 6 K_XH\dot{\pi}^3  - \dfrac{3}{\kappa}H^2,\\
\Theta &=\dfrac{1}{\kappa}H - K_X\dot{\pi}^3,\\
\Gamma &= - F_X\dot{\pi} -2F_{XX}\dot{\pi}^3 + K_\pi\dot{\pi} + K_{\pi X}\dot{\pi}^3 -9K_XH\dot{\pi}^2 - 6K_{XX}H\dot{\pi}^4,\\
\Lambda &= F_X\dot{\pi} - K_{\pi}\dot{\pi} + 3K_XH\dot{\pi}^2,
\end{align}
\end{subequations}
where $H = \frac{\dot{a}}{a}$. Varying eq.~\eqref{eq:perturb_action_gr+gal} with respect to
$\alpha$ and $\beta$,
 we obtain the following constraint equations:
\begin{subequations}
\label{eq:constraints}
\begin{align}
\dfrac{\overrightarrow{\nabla}^2\beta}{a^2}&:\;\alpha = \dfrac{1}{\Theta}\left(\dfrac{1}{\kappa}\dot{\zeta} - \dot{\chi}K_X\dot{\pi}^2 + \Lambda\chi\right), \\
\alpha&:\;\dfrac{\overrightarrow{\nabla}^2\beta}{a^2} = \dfrac{1}{\Theta}\Bigg(\dfrac{\Sigma}{\Theta}\dfrac{\dot{\zeta}}{\kappa} - \dfrac{\Sigma}{\Theta}\dot{\chi}K_X\dot{\pi}^2 + \dfrac{\Sigma}{\Theta}\Lambda\chi + 3\Theta\dot{\zeta} - \dfrac{1}{\kappa}\dfrac{\overrightarrow{\nabla}^2\zeta}{a^2} + \dfrac{\overrightarrow{\nabla}^2\chi}{a^2}K_X\dot{\pi}^2 + \Gamma\dot{\chi} \Bigg).\label{eq:constraints2}
\end{align}
\end{subequations}
As we pointed out in Sec.~\ref{sec:intro},  naively one might think that $\alpha$ and $\overrightarrow{\nabla}^2\beta$ are irrelevant in the
action \eqref{eq:perturb_action_gr+gal}, as they enter it without derivatives. However, the constraint equations \eqref{eq:constraints} show
that this is not the case: $\alpha$ and $\overrightarrow{\nabla}^2\beta$ are of the same order as derivatives of $\zeta$ and $\chi$. This is
the reason for keeping $\alpha$ and $\overrightarrow{\nabla}^2\beta$ in \eqref{eq:perturb_action_gr+gal}. Therefore, in our power counting \textit{we treat $\alpha$ and $\overrightarrow{\nabla}\beta$ as first-derivative quantities, and $\dot{\alpha}$ and $\overrightarrow{\nabla}^2\beta$ as second-derivative objects}. This property will be important in our analysis of the DPSV
approach in the next section.

Substituting the solutions to constraint equations \eqref{eq:constraints} into the action \eqref{eq:perturb_action_gr+gal} and integrating by parts, we
obtain the unconstrained quadratic  action, whose terms quadratic in derivatives are:
\begin{equation}
\label{eq:perturb_action_gr+gal_integrated}
\begin{aligned}
S^{(2)}_{gr+gal} = \int \mathrm{d}t\,\mathrm{d}^3x\,a^3\Bigg[ &\left(\dfrac{1}{\kappa^2}\dfrac{\Sigma}{\Theta^2} + \dfrac{3}{\kappa}\right)\dot{\zeta}^2 - \left( \dfrac{1}{a\cdot \kappa^2}\dfrac{\mathrm{d}}{ \mathrm{d}t}\left[\dfrac{a}{\Theta}\right] -\dfrac{1}{\kappa} \right)\dfrac{(\overrightarrow{\nabla}\zeta)^2}{a^2}\\
+&\left(\dfrac{2\Gamma}{\Theta}\dfrac{1}{\kappa} - \dfrac{2\Sigma}{\kappa\Theta^2}K_X\dot{\pi}^2\right)\dot{\zeta}\dot{\chi} + \left(\dfrac{2}{a\cdot \kappa}\dfrac{\mathrm{d}}{\mathrm{d}t}\left[K_X\dot{\pi}^2\dfrac{a}{\Theta}\right] + \dfrac{2\Lambda}{\kappa\Theta}\right)\dfrac{\overrightarrow{\nabla}\zeta\overrightarrow{\nabla}\chi}{a^2}\\
+&\left(\mathcal{A} - \dfrac{2\Gamma}{\Theta}K_X\dot{\pi}^2 + \dfrac{\Sigma}{\Theta^2}K_X^2\dot{\pi}^4\right)\dot{\chi}^2\\
-&\left(\mathcal{B} + \dfrac{1}{a}\dfrac{\mathrm{d}}{\mathrm{d}t}\left[{K_X}^2\dot{\pi}^4\dfrac{a}{\Theta}\right] + \dfrac{2\Lambda}{\Theta}K_X\dot{\pi}^2\right)\dfrac{(\overrightarrow{\nabla}\chi)^2}{a^2}\Bigg].
\end{aligned}
\end{equation}
At this point we recall that we have partially fixed the gauge in \eqref{eq:metric_perturbations}, but there remains gauge
freedom under $t \to t +\xi_0 (x)$. In terms of variables $\chi$, $\zeta$, $\alpha$ and $\beta$ we have
\begin{equation}
\label{eq:gauge_chi_alpha_beta_zeta}
\chi \to
\chi + \xi_0\dot{\pi},\quad \zeta \to \zeta + \xi_0\dfrac{\dot{a}}{a},\quad \alpha \to \alpha + \dot{\xi_0},\quad \beta \to \beta - \xi_0.
\end{equation}
The action
\eqref{eq:perturb_action_gr+gal_integrated}
can be written as follows:
\begin{equation}
\begin{aligned}
\label{eq:gauge_action_integrated}
S^{(2)}_{gr+gal} = \int \mathrm{d}t\,\mathrm{d}^3x\,a^3 \Bigg[&\dfrac{1}{\dot{\pi}^2}\left(\dfrac{1}{\kappa^2}\dfrac{\Sigma}{\Theta^2} + \dfrac{3}{\kappa}\right)\left(\dfrac{\dot{a}}{a}\dot{\chi} - \dot{\zeta}\dot{\pi}\right)^2\\
- &\dfrac{1}{\dot{\pi}^2}\left( \dfrac{1}{a\cdot \kappa^2}\dfrac{\mathrm{d}}{\mathrm{d}t}\left[\dfrac{a}{\Theta}\right] -\dfrac{1}{\kappa} \right)\left(\dfrac{\dot{a}}{a}\dfrac{\overrightarrow{\nabla}\chi}{a} - \dfrac{\overrightarrow{\nabla}\zeta}{a}\dot{\pi}\right)^2\Bigg].
\end{aligned}
\end{equation}
The combination that enters here, 
\begin{equation*}
\frac{\dot{a}}{a} \chi - \dot{\pi} \zeta \; ,
\end{equation*}
is invariant under the residual gauge transformation \eqref{eq:gauge_chi_alpha_beta_zeta}; since we are interested in high
momentum/frequency modes, we neglect  derivatives of the background when evaluating the
derivatives of this combination, so the action  \eqref{eq:gauge_action_integrated} is explicitly gauge invariant. We use
this action in Sec. \ref{sec:equiv} to study the equivalence between the DPSV and KYY approaches. 

\section{DPSV approach}
\label{sec:dpsv}
Let us discuss the DPSV approach. The main idea of the trick  is to integrate out the second
derivatives of metric perturbations  in the Galileon field equation by making use
of Einstein equations. However, as we pointed out above, there is a subtle point
that has to do with the lapse and shift perturbations. In this section we reproduce the DPSV
trick at the linearized level, but unlike in Ref.~\cite{Deffayet:2010qz} we keep terms with $\dot\alpha$ and $\overrightarrow{\nabla}^2\beta$ in the field equations. We show
that there is gauge choice under which these additional terms vanish.

The Galileon field equation for ~\eqref{eq:lagrangian} reads
\begin{equation}
\label{eq:eom_gal}
\begin{aligned}
&- 4\nabla_{\mu}\nabla_{\nu}\pi\nabla^{\mu}\pi\nabla^{\nu}\pi F_{XX} - 2\square\pi F_X + 2\square\pi K_\pi - 2\square\pi\nabla_\mu\pi\nabla^\mu\pi K_{\pi X}\\
-&4\square\pi\nabla_\mu\nabla_\nu\pi\nabla^\mu\pi\nabla^\nu\pi K_{XX} - 2\nabla_\mu\nabla^\mu\pi\nabla_\nu\nabla^\nu\pi K_X + 4\nabla_\mu\nabla_\nu\pi\nabla^\mu\pi\nabla^\nu\pi K_{\pi X}\\
+ &4\nabla_\rho\nabla_\nu\pi\nabla^\rho\nabla_\mu\pi\nabla^\mu\pi\nabla^\nu\pi K_{XX} + 2\nabla^\mu\nabla_\nu\pi\nabla_\mu\nabla^\nu\pi K_X + 2 R_{\mu\nu}\nabla^\mu\pi\nabla^\nu\pi K_X = 0,
\end{aligned}
\end{equation}
where  terms without second derivatives are omitted. The original approach~\cite{Deffayet:2010qz}
is to integrate out, with the help of Einstein equations, the second derivatives of metric
perturbations that arise from linearized $R_{\mu\nu}$.  Then the resulting linearized  equation naively
contains Galileon perturbations only. In view of the subtlety with the lapse and shift, we
now also keep terms with  $\dot{\alpha}$ and $\overrightarrow{\nabla}^2\beta$ in eq.~\eqref{eq:eom_gal}.

Let us note that quite generally, terms with second derivatives (and higher)
that can possibly arise in the linearized Galileon equation in Horndeski theory involve
\begin{equation*}
\ddot{\chi},\quad \overrightarrow{\nabla}^2\chi,\quad \ddot{\zeta},\quad \overrightarrow{\nabla}^2\zeta,\quad \dot{\alpha},\quad \ddot{\alpha},\quad\overrightarrow{\nabla}^2\alpha,\quad \overrightarrow{\nabla}^2\beta,\quad \overrightarrow{\nabla}^2\dot{\beta}
\end{equation*}
(recall that we treat $\dot{\alpha}$ and $\overrightarrow{\nabla}^2\beta$ as second-derivative quantities).
The DPSV trick eliminates $\ddot{\zeta},\,\overrightarrow{\nabla}^2\zeta,\, \ddot{\alpha},\,\overrightarrow{\nabla}^2\alpha,\,\overrightarrow{\nabla}^2\dot{\beta}$
from the Galileon field equation. As
a result, one obtains the linearized Galileon equation in terms of $\chi$, $\alpha$ and $\beta$.
This equation has to be invariant under gauge transformation \eqref{eq:gauge_chi_alpha_beta_zeta}. The only possible second order, gauge
invariant\footnote{Modulo derivatives of the background.} equation involving ${\chi}$, ${\alpha}$ and $\beta$ reads
\begin{equation}
\label{eq:vikman_action_general}
\mathcal{Q}\left(\ddot{\chi} - \dot{\alpha}\dot{\pi}\right) - \mathcal{P}\left(\overrightarrow{\nabla}^2\chi + \overrightarrow{\nabla}^2\beta\dot{\pi}\right) = 0,
\end{equation}
where $\mathcal{Q}$ and $\mathcal{P}$ depend on time through the background. 
There are two extra terms in eq.~\eqref{eq:vikman_action_general} in comparison with the original DPSV equation, namely,
the terms with $\dot{\alpha}\dot{\pi}$ and $\overrightarrow{\nabla}^2\beta\dot{\pi}$.  We now show that there is gauge in which these terms vanish.

Our key observation is that in generic Horndeski theory admitting the DPSV trick, there is a relation between lapse and shift perturbations, which is valid, to the leading order in derivatives, on solutions of the constraint equation but without gauge fixing. This relation is
\begin{equation}
\label{eq:alpha_equal_dotbeta}
\alpha = -\dot{\beta}.
\end{equation}
We give the general proof shortly, and here we notice that eq.~\eqref{eq:alpha_equal_dotbeta} is indeed satisfied in cubic theory: to the leading order in derivatives, eq.~\eqref{eq:constraints} gives $\alpha = \Theta^{-1}\left(\kappa^{-1}\dot{\zeta} - \dot{\chi}K_X\dot{\pi}^2\right)$, $\overrightarrow{\nabla}^2\beta = \Theta^{-1}\left(-\kappa^{-1}\overrightarrow{\nabla}^2\zeta + \overrightarrow{\nabla}^2\chi K_X\dot{\pi}^2\right)$. Once eq.~\eqref{eq:alpha_equal_dotbeta} holds, one can choose the comoving gauge
\begin{equation}
\label{eq:beta_equal_zero}
\beta = 0.
\end{equation}
Then, to the leading order in derivatives, $\alpha = 0$ automatically (the gauge is synchronous as well), extra terms in eq.~\eqref{eq:vikman_action_general} disappear, and we are back to the original DPSV situation, as desired.

To prove the key relation \eqref{eq:alpha_equal_dotbeta}, let us first recall 
that the second order action for perturbations~\eqref{eq:perturb_action_gr+gal} in the $\mathcal{L}_3$ theory has the following form:
\begin{equation}
\label{eq:perturb_action_general}
\begin{aligned}
S^{(2)}_{gr+gal} = \int \mathrm{d}t\,\mathrm{d}^3x\,a^3 \Bigg (
&A_1\:\dot{\zeta}^2 + A_2 \:\dfrac{(\overrightarrow{\nabla}\zeta)^2}{a^2} + A_3\: \alpha^2 + A_4\: \alpha\dfrac{\overrightarrow{\nabla}^2\beta}{a^2} + A_5\: \dot{\zeta}\dfrac{\overrightarrow{\nabla}^2\beta}{a^2} + A_6\: \alpha\dot{\zeta} +\\
&+A_7\: \alpha\dfrac{\overrightarrow{\nabla}^2\zeta}{a^2} 
+ A_8\:\alpha\dfrac{\overrightarrow{\nabla}^2\chi}{a^2} + A_9\: \dot{\chi}\dfrac{\overrightarrow{\nabla}^2\beta}{a^2} + A_{10}\:\chi\ddot{\zeta} + A_{11}\:\alpha\dot{\chi} + \\
&+A_{12}\:\chi\dfrac{\overrightarrow{\nabla}^2\beta}{a^2} 
+\mathcal{A}\dot{\chi}^2 - \mathcal{B}\dfrac{(\overrightarrow{\nabla}\chi)^2}{a^2}\Bigg),
\end{aligned}
\end{equation}
where the coefficients $A_i$ can be read off from eq.~\eqref{eq:perturb_action_gr+gal}. In fact, the quadratic action 
has the same form \eqref{eq:perturb_action_general} for general Horndeski theory with $\mathcal{L}_4$ and $\mathcal{L}_5$ included.
The only difference is that $\mathcal{L}_4$ and $\mathcal{L}_5$ subclasses introduce an additional term $A_{13}\:\chi\overrightarrow{\nabla}^2\zeta$ in the action~\eqref{eq:perturb_action_general}.
Explicit forms of the coefficients $A_i$ in general Horndeski theory are given in the Appendix.
Since the new term $\chi\overrightarrow{\nabla}^2\zeta$ involves
neither $\alpha$ nor $\beta$, it does not change the form of
constraint equations. 
Hence, the argument we give below has a general character 
and applies to complete Horndeski theory.

The coefficients $A_i$ are constrained by the fact that
the action \eqref{eq:perturb_action_general} and the linearized field equations are invariant 
under the gauge transformation \eqref{eq:gauge_chi_alpha_beta_zeta}. Let us keep those terms in the linearized equations which have the highest derivatives of the gauge parameter $\xi_0$ under the gauge transformation \eqref{eq:gauge_chi_alpha_beta_zeta}. By varying the action~\eqref{eq:perturb_action_general} with respect to $\beta$ and $\alpha$ we obtain the corresponding terms in the constraint equations:
\begin{equation}
\label{eq:alpha_and_beta}
\begin{aligned}
A_4\; &\overrightarrow{\nabla}^2\alpha + A_5\:\overrightarrow{\nabla}^2\dot{\zeta} + A_9\:\overrightarrow{\nabla}^2\dot{\chi} + \dots = 0,\\
A_4\; &\overrightarrow{\nabla}^2\beta + A_7\:\overrightarrow{\nabla}^2\zeta + A_8\:\overrightarrow{\nabla}^2\chi + \dots = 0,
\end{aligned}
\end{equation}
where dots denote terms whose gauge transformation involves lower derivatives of $\xi_0$. Invariance of these equations under the gauge transformation \eqref{eq:gauge_chi_alpha_beta_zeta} requires the following relations:
\begin{equation}
\label{eq:alpha_beta_coef}
\begin{aligned}
&A_5\: \frac{\dot{a}}{a} + A_9\:\dot{\pi} = - A_4,\\
&A_7\:  \frac{\dot{a}}{a} + A_8\: \dot{\pi} = A_4.
\end{aligned}
\end{equation}
Likewise, the equations obtained by varying the action~\eqref{eq:perturb_action_general}
with respect to $\zeta$ and  $\chi$ are
\begin{equation}
\begin{aligned}
\label{eq:zeta}
&A_7\:\overrightarrow{\nabla}^2\alpha - A_5\: \overrightarrow{\nabla}^2\dot{\beta} + \dots = 0,\\
&A_8\:\overrightarrow{\nabla}^2\alpha - A_9\: \overrightarrow{\nabla}^2\dot{\beta} + \dots = 0.
\end{aligned}
\end{equation}
Their gauge invariance implies 
\begin{equation}
\label{eq:zeta_and_chi}
A_7 = - A_5, \quad A_8 = - A_9.
\end{equation}
Now, the terms omitted in~\eqref{eq:alpha_and_beta}  actually contain fewer derivatives as compared to the terms which are written explicitly. Therefore, the relations \eqref{eq:zeta_and_chi}, together with eqs.~\eqref{eq:alpha_and_beta} give $\alpha = -\dot{\beta}$ to the leading order in derivatives, as promised.

It is worth noting that it is straightforward to
check explicitly the validity of the relations~\eqref{eq:alpha_beta_coef}
and \eqref{eq:zeta_and_chi} by making use of the expressions for
$A_i$ collected in the Appendix.

The result \eqref{eq:alpha_equal_dotbeta} is valid so long as the second order action for perturbations
has the form \eqref{eq:perturb_action_general}. This may not be the case in beyond Horndeski theory. 

In Sec. \ref{sec:DPSV_for_L4} we explicitly show that the DPSV trick works
for the $\mathcal{L}_3+\mathcal{L}_4$ case in a spatially flat FLRW background.

\section{Equivalence of the DPSV and KYY approaches}
\label{sec:equiv}

Let us now illustrate our general argument concerning DPSV by considering cubic theory. With both Galileon and
metric perturbations, linearized field equation \eqref{eq:eom_gal} reads
\begin{equation}
\label{eq:perturb_action_gal_with_gr}
-2\mathcal{A} \left(\ddot{\chi} - \dot{\alpha}\dot{\pi}\right) + 2\mathcal{B}\left(\dfrac{\overrightarrow{\nabla}^2\chi}{a^2} + \dfrac{\overrightarrow{\nabla}^2\beta}{a^2}\dot{\pi}\right) + 2R^{(1)}_{00} K_X\dot{\pi}^2 = 0,
\end{equation}
where $R^{(1)}_{00}$ is linear in perturbations. Following
the DPSV approach \cite{Deffayet:2010qz}, we obtain $R^{(1)}_{00}$ from the linearized Einstein equation,
\begin{equation}
\label{eq:riemann_zero}
R^{(1)}_{00} =\kappa\left(T_{00} - \dfrac{1}{2} T^\mu_\mu \right) = \kappa\left(-\dfrac{\overrightarrow{\nabla}^2\chi}{a^2} - 3\ddot{\chi}\right)K_X\dot{\pi}^2 + \kappa\left(3\dot{\alpha} - \dfrac{\overrightarrow{\nabla}^2\beta}{a^2}\right)K_X\dot{\pi}^3 + \dots,
\end{equation}
where the dots stand for the terms without second derivatives. Substituting this
expression for $R^{(1)}_{00}$ into eq.~\eqref{eq:perturb_action_gal_with_gr}, we get the following equation of motion
\begin{equation}
\begin{aligned}
\label{eq:vikman_action}
\left(\mathcal{A} + 3\kappa K_X^2\dot{\pi}^4\right) \left(\ddot{\chi} - \dot{\alpha}\dot{\pi}\right)
-\left(\mathcal{B} - \kappa K_X^2\dot{\pi}^4\right)\left(\dfrac{\overrightarrow{\nabla}^2\chi}{a^2} + \dfrac{\overrightarrow{\nabla}^2\beta}{a^2}\dot{\pi}\right) = 0,
\end{aligned}
\end{equation}
which does not contain $\zeta$ anymore and has precisely the form \eqref{eq:vikman_action_general}. In accordance with the above argument, the solution \eqref{eq:constraints} to constraint equations obeys \eqref{eq:alpha_equal_dotbeta}. Imposing our synchronous and comoving gauge $\alpha = \beta = 0$, we obtain the Galileon field equation 
with perturbations $\chi$ only, which is precisely the same field equation
as in \cite{Deffayet:2010qz}:
\begin{equation}
\label{eq:eq_solo_chi}
\left(\mathcal{A} + 3\kappa K_X^2\dot{\pi}^4\right) \ddot{\chi}
-\left(\mathcal{B} - \kappa K_X^2\dot{\pi}^4\right)\dfrac{\overrightarrow{\nabla}^2\chi}{a^2} = 0.
\end{equation}
Hence, omitting terms  $\dot{\alpha}$ and $\overrightarrow{\nabla}^2\beta$ in \eqref{eq:vikman_action} is equivalent to choosing the gauge $\beta = 0$.

Field equation \eqref{eq:eq_solo_chi} can be derived from the following action:
\begin{equation}
\label{eq:vikman_action_integrated}
S^{(2)}_{gal} = \int \mathrm{d}t\,\mathrm{d}^3x\,a^3\left[\left(\mathcal{A} + 3\kappa K_X^2\dot{\pi}^4\right) \dot{\chi}^2 - \left(\mathcal{B} - \kappa K_X^2\dot{\pi}^4\right)\dfrac{(\overrightarrow{\nabla}\chi)^2}{a^2}\right],
\end{equation}
which we now use to show the explicit equivalence of the KYY and DPSV
approaches.

Making use of eq.~\eqref{eq:notations_in_action}, we cast the action \eqref{eq:vikman_action_integrated} in the following form
\begin{equation}
\label{eq:action_dpsv}
S^{(2)}_{gr+gal} = \int \mathrm{d}t\,\mathrm{d}^3x\,a^3 \cdot\dfrac{\kappa^2\Theta^2}{\dot{\pi}^2}\left[\left(\dfrac{1}{\kappa^2}\dfrac{\Sigma}{\Theta^2} + \dfrac{3}{\kappa}\right)\dot{\chi}^2 - \left( \dfrac{1}{a\cdot \kappa^2}\dfrac{\mathrm{d}}{ \mathrm{d}t}\left[\dfrac{a}{\Theta}\right] -\dfrac{1}{\kappa} \right)\dfrac{(\overrightarrow{\nabla}\chi)^2}{a^2}\right].
\end{equation}
This action is obtained from the action \eqref{eq:gauge_action_integrated} by setting
\begin{equation}
\label{eq:zeta_chi}
\zeta = \kappa K_X\dot{\pi}^2\chi.
\end{equation}
The latter relation is, in fact, a consequence of the constraint equations \eqref{eq:constraints}
in the gauge $\alpha=\beta=0$ (to the leading order in derivatives). This demonstrates 
the consistency of the entire approach.

On the other hand, the results of the KYY approach correspond to choosing $\chi=0$  in action \eqref{eq:gauge_action_integrated} (unitary gauge):
\begin{equation}
\label{eq:action_kyy}
S^{(2)}_{gr+gal} = \int \mathrm{d}t\,\mathrm{d}^3x\,a^3 \left[\left(\dfrac{1}{\kappa^2}\dfrac{\Sigma}{\Theta^2} + \dfrac{3}{\kappa}\right)\dot{\zeta}^2 - \left( \dfrac{1}{a\cdot \kappa^2}\dfrac{\mathrm{d}}{\mathrm{d}t}\left[\dfrac{a}{\Theta}\right] -\dfrac{1}{\kappa} \right)\dfrac{(\overrightarrow{\nabla}\zeta)^2}{a^2}\right].
\end{equation}
This expression has been derived for the cubic Galileon \eqref{eq:lagrangian} in Refs. \cite{Kobayashi:2010cm, Kobayashi:2011nu}. Modulo field redefinition and non-derivative terms, the actions \eqref{eq:action_dpsv} and \eqref{eq:action_kyy} coincide. So, we see that the results of both approaches are equivalent to each other,  inasmuch as they
can be derived from the gauge invariant action \eqref{eq:gauge_action_integrated} by choosing a particular gauge.

\section{DPSV trick for $\mathcal{L}_4$}
\label{sec:DPSV_for_L4}
As shown in the previous section, the DPSV approach corresponds to a specific choice
of gauge in the invariant action \eqref{eq:gauge_action_integrated}. A natural question that arises is whether the same
trick is applicable to a Horndeski theory that includes $\mathcal{L}_3$ and $\mathcal{L}_4$:
\begin{equation*}
\begin{aligned}
&\mathcal{L}_3 = F(\pi,X) + K(\pi,X)\square\pi,\\
&\mathcal{L}_4= - G_4(\pi,X)R + 2G_{4X}(\pi,X)\left[\left(\Box\pi\right)^2-\nabla_\mu\nabla_\nu\pi \; \nabla^\mu\nabla^\nu\pi \right].
\end{aligned}
\end{equation*}
The case considered in previous sections corresponds to $G_4(\pi,X)=\frac{1}{2\kappa}$. Including the $\mathcal{L}_4$ term
in the action leads to the appearance of extra terms with second derivatives of the metric in the
Galileon field equation. Together with the term due to $\mathcal{L}_3$, these are
\begin{equation}
\label{galileon_eq_L4}
\begin{aligned}
&2  R_{\mu\nu}  \nabla^{\mu} \pi \nabla^{\nu} \pi \: K_{X}+
2 R \nabla_{\mu} \nabla^{\mu} \pi \: G_{4X} - 
4 R_{\mu\nu}  \nabla^{\mu} \nabla^{\nu} \pi \: G_{4X} +\\+
&4 R \nabla^{\mu}\pi \nabla_{\mu}  \nabla_{\nu} \pi \nabla^{\nu}\pi \: G_{4XX} -
16 R_{\mu\nu} \nabla^{\mu}\pi \nabla^{\nu}\nabla_{\rho}\pi \nabla^{\rho}\pi \: G_{4XX}+
8 R_{\mu\nu}  \nabla^{\mu} \pi \nabla^{\nu} \pi \nabla_{\rho}\nabla^{\rho}\pi \: G_{4XX} -\\-
& R\: G_{4\pi} + 2 R \nabla_{\mu}\pi \nabla^{\mu}\pi \: G_{4\pi X} -
8 R_{\mu\nu} \nabla^{\mu}\pi \nabla^{\nu}\pi \: G_{4\pi X} -
8 R_{\mu\nu\rho\sigma}  \nabla^{\mu} \pi \nabla^{\rho} \pi \nabla^{\nu}\nabla^{\sigma}\pi \:G_{4XX}+ \dots = 0,
\end{aligned}
\end{equation}
where the dots stand for terms without second derivatives of the metric. Like in Sec. \ref{sec:dpsv}, the Galileon field equation contains second derivatives of both the metric and Galileon. One would like to eliminate the second derivatives of the metric by making use of the Einstein equations. The  latter have the following form
\begin{equation}
\label{Einstein_L4}
\begin{aligned}
& 2 G_{\mu\nu} \:G_{4}  + 
2 R \nabla_{\mu} \pi \nabla_{\nu} \pi \: G_{4X} -
4 R_{\nu\rho} \nabla_{\mu}\pi \nabla^{\rho}\pi \: G_{4X} -\\-
& 4 R_{\mu\rho} \nabla_{\nu}\pi \nabla^{\rho}\pi \: G_{4X} +
4 g_{\mu\nu} R_{\rho\sigma} \nabla^{\rho}\pi \nabla^{\sigma}\pi \: G_{4X} - 
4 R_{\mu\rho\nu\sigma} \nabla^{\rho}\pi \nabla^{\sigma} \pi \: G_{4X} + \dots = 0,
\end{aligned}
\end{equation}
where dots again stand for terms without second derivatives of the metric.

The issue is now whether or not the linearized equations \eqref{Einstein_L4} enable one to get rid of the second derivatives of the metric in linearized equation \eqref{galileon_eq_L4}. For a general background the answer is negative: the second derivative term in the linearized expression \eqref{galileon_eq_L4} is not a linear combination of the second derivative terms in the linearized expressions \eqref{Einstein_L4}. We have checked that explicitly by calculating the rank of the relevant matrix, with background quantities like $\nabla_\mu\pi$, $\nabla_\mu\nabla_\nu\pi$, etc. taking random numerical values. So, the DPSV trick does not work for $\mathcal{L}_4$ theory, unlike in the $\mathcal{L}_3$ case.

However, if one assumes that the background Galileon field is homogeneous, i.e.,
$\pi_c=\pi(t)$, and the background geometry is of the spatially flat FLRW type, the trick for $\mathcal{L}_4$ comes back.
The only Galileon derivatives for the homogeneous background are $\dot{\pi}$, $\ddot{\pi}$
and $\nabla_i \nabla_j \pi = g_{ij} \dot{\pi} H$.\footnote{In what follows Latin indices take values $1$, $2$, $3$.} Therefore, Galileon field equation \eqref{galileon_eq_L4} involves a reduced number of non-trivial components of $R_{\mu\nu\rho\sigma}$. Indeed, the linearized Galileon field equation simplifies to
\begin{equation}
\label{gal_eq_expanded}
\begin{aligned}
&g^{ij} g^{mn} R^{(1)}_{imjn} \left(-G_{4\pi} + 2 G_{4X} H \dot{\pi} + 2 G_{4X} \ddot{\pi} + 2 \dot{\pi}^2 (G_{4\pi X} + 2 G_{4XX} \ddot{\pi}) \right) -\\
&-2 g^{ij}  R^{(1)}_{i0j0} \left(G_{4\pi} - 4G_{4X} H \dot{\pi} + (2 G_{4\pi X} - K_{X})\dot{\pi}^2 - 8 G_{4XX} H \dot{\pi}^3 \right) + Z = 0,
\end{aligned}
\end{equation}
where $Z$ denotes terms with second derivatives of the Galileon perturbations, but without second derivatives of the metric, and in view of the general discussion in Sec. \ref{sec:dpsv}, we choose the comoving and synchronous gauge $\alpha = \beta = 0$ from the very beginning. From eq.~(\ref{gal_eq_expanded}) we see that in order to obtain the equation that is free of second derivatives of metric
perturbations, we need two structures expressed in terms of Galileon perturbations, namely
$g^{ij} g^{mn} R^{(1)}_{imjn}$ and $g^{ij}  R^{(1)}_{i0j0}$. The same structures appear in the $00-\mbox{component}$ and  trace (i.e.
$G_{\mu\nu}g^{\mu\nu}$) of the linearized Einstein equations:
\begin{subequations}
\begin{align}
\label{einst_00}
&g^{ij} g^{mn} R^{(1)}_{imjn} (G_{4} - 2 G_{4X} \dot{\pi}^2) = Y_{00} ,\\
\label{einst_trace}
& 4 g^{ij}  R^{(1)}_{i0j0} (G_{4} - 2 G_{4X} \dot{\pi}^2) + 2 g^{ij} g^{mn} R^{(1)}_{imjn} (G_{4} -  G_{4X} \dot{\pi}^2) = g^{\mu\nu} Y_{\mu\nu},
\end{align}
\end{subequations}
where $Y_{\mu\nu}$ and its component $Y_{00}$ again do not contain the second derivatives of the metric. Expressing $g^{ij} g^{mn} R^{(1)}_{imjn}$
and $g^{ij}  R^{(1)}_{i0j0}$ through eqs.~(\ref{einst_00}) and (\ref{einst_trace}), we obtain the linearized Galileon field equation (\ref{gal_eq_expanded}) 
without second derivatives of metric perturbations:
\begin{equation}
\label{gal_eq_after_trick}
\begin{aligned}
&\frac{Y_{00}}{ \left(G_{4} - 2 G_{4X} \dot{\pi}^2 \right)} \left(-G_{4\pi} + 2 G_{4X} H \dot{\pi} + 2 G_{4X} \ddot{\pi} + 2 \dot{\pi}^2 (G_{4\pi X} + 2 G_{4XX} \ddot{\pi}) \right) -\\
&-2 \left[ \frac{g^{\mu\nu} Y_{\mu\nu}}{4 \left(G_{4} -  2G_{4X} \dot{\pi}^2 \right)} - \frac{Y_{00} \left(G_{4} -  G_{4X} \dot{\pi}^2 \right)}{2 \left(G_{4} - 2 G_{4X} \dot{\pi}^2 \right)^2} \right] \left(G_{4\pi} - 4G_{4X} H \dot{\pi} \right. +\\
&\left.+(2 G_{4\pi X} - K_{X})\dot{\pi}^2 - 8 G_{4XX} H \dot{\pi}^3\right) + Z = 0,
\end{aligned}
\end{equation}
Making use of eq.~(\ref{gal_eq_after_trick}) we reconstruct the analogue of \eqref{eq:vikman_action_integrated} for the $\mathcal{L}_3+\mathcal{L}_4$ theory. Straightforward calculations lead to the following result:
\begin{equation}
\label{eq:vikman_action_integrated_L4}
S^{(2)}_{gal} = \int \mathrm{d}t\,\mathrm{d}^3x\,a^3\left[\tilde{\mathcal{A}} \dot{\chi}^2 - \tilde{\mathcal{B}}\dfrac{(\overrightarrow{\nabla}\chi)^2}{a^2}\right],
\end{equation}
where
\begin{equation*}
\begin{aligned}
\tilde{\mathcal{A}} =\dfrac{\tilde{\Sigma}\mathcal{G_T}+3 \tilde{\Theta}^2}{\mathcal{G_T} \dot{\pi}^2},\quad \tilde{\mathcal{B}} = \dfrac{\tilde{\Theta}^2}{\mathcal{G_T}^2 \dot{\pi}^2}\left(\dfrac{1}{a}\dfrac{\mathrm{d}}{\mathrm{d}t}\left[\dfrac{a\mathcal{G_T}^2}{\tilde{\Theta}}\right]-2 G_4\right),
\end{aligned}
\end{equation*}
and
\begin{subequations}
	\begin{align*}
	&\mathcal{G_T}=2G_4-4G_{4X}\dot{\pi}^2,\\
	&\tilde{\Sigma} = \mathcal{A} \dot{\pi}^2 + 6K_XH\dot{\pi}^3 -6G_4H^2+42G_{4X}H^2 \dot{\pi}^2+96G_{4XX}H^2 \dot{\pi}^4+
	\\\nonumber&+24G_{4XXX}H^2 \dot{\pi}^6
	-6G_{4\pi}H\dot{\pi}-30G_{4\pi X}H \dot{\pi}^3-12G_{4\pi XX}H \dot{\pi}^5,\\
	&\tilde{\Theta}=-K_X\dot{\pi}^3+2G_4H-8G_{4X}H\dot{\pi}^2-8G_{4XX}H\dot{\pi}^4+G_{4\pi}\dot{\pi}+2G_{4\pi X}\dot{\pi}^3.
	\end{align*}
\end{subequations}
On the other hand, the gauge invariant  action for perturbations in the $\mathcal{L}_3+\mathcal{L}_4$ case reads
\begin{equation*}
\begin{aligned}
S^{(2)}_{gr+gal} = \int \mathrm{d}t\,\mathrm{d}^3x\,a^3 \Bigg[&\dfrac{1}{\dot{\pi}^2}\left(\dfrac{\tilde{\Sigma}\mathcal{G_T}^2}{\tilde{\Theta}^2} + 3\mathcal{G_T}\right)\left(\dfrac{\dot{a}}{a}\dot{\chi} - \dot{\zeta}\dot{\pi}\right)^2\\
- &\dfrac{1}{\dot{\pi}^2}\left( \dfrac{1}{a}\dfrac{\mathrm{d}}{\mathrm{d}t}\left[\dfrac{a\mathcal{G_T}^2}{\tilde{\Theta}}\right] -2G_4 \right)\left(\dfrac{\dot{a}}{a}\dfrac{\overrightarrow{\nabla}\chi}{a} - \dfrac{\overrightarrow{\nabla}\zeta}{a}\dot{\pi}\right)^2\Bigg].
\end{aligned}
\end{equation*}
In the unitary gauge ($\chi=0$) this action is given in \cite{Kobayashi:2011nu}, 
while in the gauge $\alpha=\beta=0$ we have
 $\zeta = \chi \; \frac{H \mathcal{G_T} - \tilde\Theta}{ \mathcal{G_T} \dot{\pi}^2}$,
and this action coincides with (\ref{eq:vikman_action_integrated_L4}) obtained via the extended DPSV trick. As in the $\mathcal{L}_3$ case, we observe that integrating out the second derivatives of the metric corresponds to choosing a particular gauge.

\section*{Acknowledgements}
The authors are indebted to A. Vikman for useful comments.
This work has been supported by Russian Science Foundation Grant No. 
14-22-00161.

\section*{Appendix}
In this section we give explicit expressions  for coefficients $A_i$
entering
the action \eqref{eq:perturb_action_general}, for the general Horndeski theory with the Lagrangian
\begin{equation*}
	\begin{aligned}
	&\mathcal{L} = \mathcal{L}_3 + \mathcal{L}_4 + \mathcal{L}_5,\\
	&\mathcal{L}_3 = F(\pi,X) + K(\pi,X)\square\pi,\\
&\mathcal{L}_4= - G_4(\pi,X)R + 2G_{4X}(\pi,X)\left[\left(\Box\pi\right)^2-\nabla_\mu\nabla_\nu\pi \; \nabla^\mu\nabla^\nu\pi \right],\\
	&\mathcal{L}_5=G_5(\pi,X)G^{\mu\nu}\nabla_{\mu}\nabla_{\nu} \pi+\\
	&+\frac{1}{3}G_{5X}\left[\left(\Box\pi\right)^3-3\Box\pi \nabla_{\mu}\nabla_{\nu} \pi \nabla^{\mu}\nabla^{\nu} \pi +2\nabla_{\mu}\nabla_{\nu} \pi \nabla^{\mu}\nabla^{\rho} \pi \nabla_{\rho}\nabla^{\nu} \pi \right].
	\end{aligned}
\end{equation*}
These expressions are
\begin{subequations}
\begin{align*}
&A_1=3\left[-2G_4+4G_{4X}\dot{\pi}^2-G_{5\pi}\dot{\pi}^2+2HG_{5X}\dot{\pi}^3\right],\\
&A_2=2G_4-2G_{5X}\dot{\pi}^2\ddot{\pi}-G_{5\pi}\dot{\pi}^2,\\
&A_3=F_X\dot{\pi}^2+2F_{XX}\dot{\pi}^4+12HK_X\dot{\pi}^3+6HK_{XX}\dot{\pi}^5-K_{\pi}\dot{\pi}^2-K_{\pi X}\dot{\pi}^4\\
\nonumber&-6H^2G_4+42H^2G_{4X}\dot{\pi}^2+96H^2G_{4XX}\dot{\pi}^4+24H^2G_{4XXX}\dot{\pi}^6\\
\nonumber&-6HG_{4\pi}\dot{\pi}-30HG_{4\pi X}\dot{\pi}^3-12HG_{4\pi XX}\dot{\pi}^5+30H^3G_{5X}\dot{\pi}^3\\
\nonumber&+26H^3G_{5XX}\dot{\pi}^5+4H^3G_{5XXX}\dot{\pi}^7-18H^2G_{5\pi}\dot{\pi}^2-27H^2G_{5\pi X}\dot{\pi}^4\\
\nonumber&-6H^2G_{5\pi XX}\dot{\pi}^6,\\
&A_4=2\big[K_X\dot{\pi}^3-2G_4H+8HG_{4X}\dot{\pi}^2+8HG_{4XX}\dot{\pi}^4-G_{4\pi}\dot{\pi}-2G_{4\pi X}\dot{\pi}^3\\ 
\nonumber&+5H^2G_{5X}\dot{\pi}^3+2H^2G_{5XX}\dot{\pi}^5-3HG_{5\pi}\dot{\pi}^2-2HG_{5\pi X}\dot{\pi}^4\big],\\
&A_5=-\dfrac{2}{3}A_1,\\
&A_6=-3A_4,\\
&A_7=\dfrac{2}{3}A_1 = -A_5,\\
&A_8=2\big[K_X\dot{\pi}^2-G_{4\pi}-2G_{4\pi X}\dot{\pi}^2+4 H\dot{\pi}G_{4X}+8HG_{4XX}\dot{\pi}^3-2HG_{5\pi}\dot{\pi}\\
&-2HG_{5\pi X}\dot{\pi}^3+3 H^2G_{5X}\dot{\pi}^2+2 H^2G_{5XX}\dot{\pi}^4\big],\\
&A_9=-A_8,\\
&A_{10}=-3A_8,\\
\end{align*}
\end{subequations}
\begin{subequations}
\begin{align*}
&A_{11}=2\big[-F_X\dot{\pi}-2F_{XX}\dot{\pi}^3+K_\pi\dot{\pi}-6HK_{XX}\dot{\pi}^4-9HK_X\dot{\pi}^2+K_{\pi X}\dot{\pi}^3\\
&+3HG_{4\pi}+24HG_{4\pi X}\dot{\pi}^2+12H G_{4\pi XX}\dot{\pi}^4-18H^2G_{4X}\dot{\pi}-72H^2G_{4XX}\dot{\pi}^3\\
&-24H^2G_{4XXX}\dot{\pi}^5+9H^2G_{5\pi}\dot{\pi}+21H^2G_{5\pi X}\dot{\pi}^3+6H^2G_{5\pi XX}\dot{\pi}^5\\
&-15H^3G_{5X}\dot{\pi}^2-
20H^3G_{5XX}\dot{\pi}^4-4H^3G_{5XXX}\dot{\pi}^6\big],\\
&A_{12}=2\big[F_X\dot{\pi}-K_\pi\dot{\pi}+3HK_X\dot{\pi}^2-HG_{4\pi}+G_{4\pi\pi}\dot{\pi}-10HG_{4\pi X}X+6H^2G_{4X}\dot{\pi}\\
&+12H^2G_{4XX}\dot{\pi}^3-3H^2G_{5\pi}\dot{\pi}+HG_{5\pi\pi}\dot{\pi}^2-4H^2G_{5\pi X}\dot{\pi}^3+3H^3G_{5X}\dot{\pi}^2\\
&+2H^3G_{5XX}\dot{\pi}^4\big],\\
&A_{13}= 2\big[4HG_{4X}\dot{\pi} + 4G_{4X}\ddot{\pi} + 8 G_{4XX}\dot{\pi}^2\ddot{\pi} - 2G_{4\pi} +4 G_{4\pi X}\dot{\pi}^2 + 2 H^2G_{5X}\dot{\pi}^2\\
&+2 \dot{H}G_{5X}\dot{\pi}^2+4 H G_{5X}\dot{\pi}\ddot{\pi}+4HG_{5XX}\dot{\pi}^3\ddot{\pi} -2HG_{5\pi}\dot{\pi}-2G_{5\pi}\ddot{\pi}+2HG_{5\pi X}\dot{\pi}^3\\
&-2G_{5\pi X}\dot{\pi}^2\ddot{\pi}-G_{5\pi\pi}\dot{\pi}^2\big].
\end{align*}
\end{subequations}
We see that eqs.~\eqref{eq:zeta_and_chi}, which we obtained on the basis of gauge invariance, are satisfied explicitly. It is also straightforward
to check explicitly that eqs.~\eqref{eq:alpha_beta_coef} are satisfied as well.

\end{document}